\documentclass[sts, preprint]{imsart}

\usepackage{amsthm,amsmath,natbib}
\RequirePackage[colorlinks,citecolor=blue,urlcolor=blue]{hyperref}
\startlocaldefs
\endlocaldefs

\begin{document}

\begin{frontmatter}

% "Title of the paper"
\title{Benchmarking in cluster analysis: A white paper}
\runtitle{Benchmarking in cluster analysis}

% indicate corresponding author with \corref{}
% \author{\fnms{John} \snm{Smith}\corref{}\ead[label=e1]{smith@foo.com}\thanksref{t1}}
% \thankstext{t1}{Thanks to somebody} 
% \address{line 1\\ line 2\\ printead{e1}}
% \affiliation{Some University}

\begin{aug}
\author{\fnms{Iven} \snm{Van Mechelen}\corref{}\ead[label=e1]{iven.vanmechelen@kuleuven.be}},
\author{\fnms{Anne-Laure} \snm{Boulesteix}\ead[label=e2]{boulesteix@ibe.med.uni-muenchen.de}},
\author{\fnms{Rainer} \snm{Dangl}\ead[label=e3]{rainer.dangl@gmail.com}},
\author{\fnms{Nema} \snm{Dean}\ead[label=e4]{nema.dean@glasgow.ac.uk}},
\author{\fnms{Isabelle} \snm{Guyon}\ead[label=e5]{isabelle@clopinet.com}},
\author{\fnms{Christian} \snm{Hennig}\ead[label=e6]{c.hennig@ucl.ac.uk}},
\author{\fnms{Friedrich} \snm{Leisch}\ead[label=e7]{friedrich.leisch@boku.ac.at}},
\and
\author{\fnms{Douglas} \snm{Steinley}\ead[label=e8]{steinleyd@missouri.edu}\thanksref{t1}}
\thankstext{t1}{The work on this paper has been supported in part by the Interuniversity Attraction Poles program of the Belgian government (grant IAP P7/06 to Iven Van Mechelen), by the Research Fund of KU Leuven (grant GOA/2015/03 to Iven Van Mechelen), by the German Research Foundation (grant BO3139/2-3 to Anne-Laure Boulesteix), and by the Engineering and Physical Sciences Research Council (grant EP/K033972/1 to Christian Hennig).}

\runauthor{Van Mechelen et al.}

%\affiliation{; ; ; ; ; ; }
\address{Iven Van Mechelen: University of Leuven, Belgium \printead{e1}, Anne-Laure Boulesteix: Ludwig-Maximilians-University Munich, Germany \printead{e2}, Rainer Dangl: University of Natural Resources and Life Sciences, Vienna,  Austria \printead{e3}, Nema Dean: University of Glasgow, Glasgow, UK \printead{e4}, Isabelle Guyon: University of Paris-Saclay, France \printead{e5}, Christian Hennig: University College London, UK \printead{e6}, Friedrich Leisch: University of Natural Resources and Life Sciences, Vienna, Austria  \printead{e7}, Douglas Steinley: University of Missouri, USA \printead{e8}}
\end{aug}

\begin{abstract}
{\bf A revised version is now published!} Please cite (and read; it's open access)
\\~\\
 Van Mechelen, I., Boulesteix, A.-L., Dangl, R., Dean, N., Hennig, C., Leisch, F., Steinley, D., \& Warrens, M. J. (2023). {\it A white paper on good research practices in benchmarking: The case of cluster analysis.} WIREs Data Mining and Knowledge Discovery, e1511.\\ \verb|https://doi.org/10.1002/widm.1511|
~\\~\\
To achieve scientific progress in terms of building a cumulative body of knowledge, careful attention to benchmarking must be of the utmost importance. This means that proposals of new methods of data pre-processing, new data-analytic techniques, and new methods of output post-processing, should be extensively and carefully compared with existing alternatives, and that existing methods should be subjected to neutral comparison studies. To date, benchmarking and recommendations for benchmarking have been frequently seen in the context of supervised learning. Unfortunately, there has been a dearth of guidelines for benchmarking in an unsupervised setting, with the area of clustering as an important subdomain. To address this problem, discussion is given to the theoretical conceptual underpinnings of benchmarking in the field of cluster analysis by means of simulated as well as empirical data. Subsequently, the practicalities of how to address benchmarking questions in clustering are dealt with, and foundational recommendations are made.
\end{abstract}

% 62H30 	Classification and discrimination; cluster analysis

\end{frontmatter}

\section{Introduction}\label{sec:intro}
In any data-analytic process, many choices are to be made on the level of the pre-processing of the input of the process (viz., the actual data), the data analysis in the narrow sense, and the post-processing of the data-analytic output. Apart from many alternatives being available for all choices in question, new methods are continually being developed on all three levels.

	An obvious question at this point is which choice alternatives are optimal in some respect, which comes down to a question of comparative evaluation. This is an important question for developers of methods and methodological researchers. Indeed, when new methods are proposed, one may wish these to be justified in terms of an in-depth comparison with their best predecessors, with good science necessarily being cumulative in nature. Apart from that, the question of comparative evaluation is also of utmost importance to the data-analytic practitioner who may desperately need aid in making good choices in the analysis of applications.

	We will further define addressing questions of comparative evaluation as an exercise in benchmarking. There is a clear benchmarking tradition in many subdomains of data analysis. In particular, there is a strong such tradition in the domain of statistical learning, and especially in the supervised part of this domain (which includes (supervised) classification as an important special case) \citep[see e.g.,][]{eugster11, stone74}.

	However, the situation is rather different in the domain of unsupervised learning, including, in particular, (unsupervised) cluster analysis. To be sure, the clustering domain has witnessed a few seminal benchmarking attempts \citep{milligan80, milligan85, milligan85a, milligan88}, followed by work from several other authors \citep[see, e.g.,][]{steinley03}. However, in general, in this domain there is much less of a benchmarking tradition. This is, for instance, evidenced by the fact that very often new methods are proposed without a sound comparison with predecessors, which obviously seriously hampers a cumulative building of knowledge. Also, within the clustering domain there is a dearth of recommendations and guidelines for benchmarking. The latter goes with the fact that, in the domain of cluster analysis, benchmarking is less straightforward, and comes with quite a few challenges. A major reason for this is that in supervised classification, classes and class membership are known a priori; hence, this naturally leads to a concrete metric for evaluating performance \citep[viz., the error rate or proportion of objects misclassified: see, e.g.,][]{schiavo00}. In contrast, in an unsupervised clustering context, classes and class memberships are unknown, and must be ``discovered'' by the clustering algorithm. This results in quite a few obstacles with regard to determining how well a specific approach is performing. (Incidentally, one might hope to cope with these obstacles by using data sets with known clusters for benchmarking in unsupervised clustering, but this is not without its problems: see \ref{sec:evaluationissues}.)
	
	To some extent, mathematical theories can serve and (at least indirectly) compare the performance of different cluster analysis methods. Examples of such theories include consistency and optimality theories for probability models for clustering, and axiomatic approaches to characterize clustering methods. However, in reality the formal assumptions of such theories (insofar they exist) are usually not fulfilled, and their relevance for practical applications may be controversial. Benchmarking based on the analysis of data sets will always at the very least deliver useful complementary information.

	The goal of the current paper is to present a discussion of the most fundamental conceptual underpinnings of data-based benchmarking in the cluster analysis domain, along with a set of foundational guidelines. It is hoped that this endeavor will foster scientific debate on benchmarking in the area, and that this will subsequently lead to improved scientific practice.

	The remainder of this paper is structured as follows: Section \ref{sec:concept} discusses the theoretical conceptual underpinnings of benchmarking in the clustering area; Section \ref{sec:practicalities} continues with a discussion of the practicalities of how to address benchmarking questions in clustering along with foundational guidelines; Section \ref{sec:examples} presents a handful of illustrative examples. The paper closes with some concluding remarks (Section \ref{sec:conclusion}) and a glossary with summary definitions of a few key terms (Section \ref{sec:glossary}).
	
\section{Benchmarking in clustering: Conceptual underpinnings}\label{sec:concept}
In this section, we outline a conceptual framework to structure the problem of benchmarking in cluster analysis in terms of some major fundamental distinctions. Successively, we present a framework for: (1) the choice alternatives that are to be compared in benchmarking studies of clustering, (2) the data that are to be used in such studies, (3) the quality or performance criteria the studies may focus on, and (4) the type of answers the studies may lead to.

\subsection{Choice alternatives to be compared in benchmarking studies of clustering}\label{sec:choice}
The choice alternatives that are to be compared may concern a broad range of aspects of the data-analytic process of clustering. Below we will present a structured overview of the different steps in the clustering process. This will yield fundamental conceptual underpinnings for understanding both the different possible methods that may be compared in benchmarking studies and the different possible evaluation criteria that may be used in this comparison.
\subsubsection{Starting point of data-analytic process of clustering - Research goal and data:}\label{sec:start} 

\vspace*{0.5cm}The starting of any data-analytic process comprises, on the one hand, the research goal, and, on the other hand, the data by means of which one may wish to achieve this goal. The research goal of any clustering process is to learn or induce an unknown clustering from the data, with the clustering being defined in terms of cluster memberships (i.e., the clusters' \textit{extension}) and/or the clusters' characteristics, definitions, and membership rules (i.e., the clusters' \textit{intension}) \citep{leibniz04}. The underlying aim at this point can be the search for the ``true'' clustering \citep{hennig15}, which can be linked to the Socratic target of ``carving nature at its joints'' \citep{plato}, and which can be qualified as fundamental or ``realistic''. Alternatively, the underlying goal could be more pragmatic or ``constructive'' \citep{hennig15a} in nature, with the ideal clustering being optimal in terms of some extrinsic goal such as information compression (with the compressed information optionally to be used in some subsequent data analysis), communication, or providing a suitable basis for some kind of action. Apart from the realistic-constructive distinction, one should also consider the question whether induction of a clustering is the only research goal at issue, or rather if it goes with another goal. Examples include the combination of the induction of a clustering with a goal of prediction \citep[as, e.g., in mixtures of mixed models: ][]{celeux05}, of dimension reduction \citep[as, e.g., in mixtures of factor analyzers:][]{mclachlan03}, or of process modeling \citep[as, e.g., in mixtures of hidden Markov models:][]{smyth97}. In the latter cases, the primary concern may be either in the clustering or in the other goal, even for one and the same clustering method. 

	The second major constituent of the starting point of the data-analytic process is the type of the data by means of which one would like to address the research goal. In terms of data structure, two basic types of data are most frequently encountered in the clustering domain: simple object by variable data and simple object by object proximity (similarity or dissimilarity) data with unstructured object or variable data modes. Apart from these, quite a few more complex data structures may be encountered as well, including graph or network data, multiway data, multiblock data (including hierarchical or multilevel data as an important special case), and data with structured object or variable data modes (e.g., time or space) such as sequence or functional data. In addition to these important distinctions with regard to the structure of the data, an important remark has to be made: Leaving aside semi-supervised cases, data for a cluster analysis normally do not include a priori known cluster memberships. If such information were nonetheless available, it would be kept apart from the ``internal'' part of the data that is to be used in the cluster analysis, and could optionally be used afterwards as ``external'' information to validate clustering results.

	Given this starting point, we can now move on to the three steps of the data-analytic process as already listed in the introduction: (1) the pre-processing of the data, (2) the data analysis in the narrow sense (i.e., leaving aside the other two steps), and (3) the post-processing of the output of the second step. Below we will successively outline the most important conceptual distinctions for each of these three steps.

\subsubsection{Level 1 - Pre-processing of the data:}\label{sec:level1}

Pre-processing can take quite many different forms that vary in the degree that the original data are transformed. One of the more modest forms of pre-processing involves various kinds of data selection. Such a selection can refer to the experimental units or objects with, for example, a removal of outliers prior to the data analysis in the narrow sense. Alternatively, selection can also refer to the variables included in the data set. Otherwise, data selection can take place both prior to and during the actual data analysis, with only the former qualifying as data pre-processing.

	Another relatively more limited type of preprocessing is that of transformations of variables (or even entire data blocks) prior to the data analysis in the narrow sense. Classic examples that have been frequently addressed in the clustering domain are various combinations of centering and scaling \citep[e.g.,][]{milligan88, steinley04}. Once again, data transformations as a way of data-preprocessing have to be distinguished from data transformations that are part of some optimization in the data-analytic process in the narrow sense \citep[as in, e.g., various methods that involve some kind of optimal scaling -- for an illustration, see][]{vanbuuren89}.

	The most comprehensive types of preprocessing imply a conversion of the data structure into a different type of structure prior to the data analysis in the narrow sense. A classic instance of this is the conversion of object by variable data into object by object proximities (for which an unlimited wealth of proximity measures could be invoked -- see, e.g., \cite{gower86} --, and which therefore implies a major challenge on the level of choices to be made). Note that this type of conversion implies that clustering methods based on proximity data can be compared in benchmarking with methods based on object by variable data (or any other type of data that can be converted into proximities). 

\subsubsection{Level 2 - Data analysis in the narrow sense:}\label{sec:level2}
Aspects of the data analysis in the narrow sense include the nature of the clusters aimed at, the criterion or objective function that is to be optimized, and the choice of a suitable algorithm or computational procedure. Below we will successively discuss these three aspects. \\

\noindent\textit{Nature of the clusters} \\

Perhaps the most fundamental aspect of the data analysis in the narrow sense is the nature of the clustering aimed at. This nature can be typified in terms of set-theoretical characteristics of the membership of the clusters of interest (i.e., their extension) and of organizing principles on the level of the clusters' definition, associated features or attributes (i.e., their intension).

	The following five questions can be helpful to typify the nature of the set-theoretical characteristics of the clusters' \textbf{extension}: 
\begin{enumerate}
\item \textit{Nature of the elements of the clusters.} The elements of the clusters can be objects, variables, or some other type of entities involved in the data (e.g., sources). Alternatively, in case of data that involve two or more types of entities (such as object by variable or source by object by variable data), the elements to be clustered could be $n$-tuples such as ordered pairs (of, e.g., an object and a variable) in two-mode, bi- or co-clustering \citep[see, further, e.g.,][]{hartigan75, vanmechelen04}. 
\item \textit{Crisp or fuzzy membership?} The most straightforward form of cluster membership is crisp in nature, with elements either not belonging (membership=0) or belonging (membership=1) to a cluster under study. As an alternative, one may consider graded forms of membership (with membership values varying from 0 to 1), which optionally could be formalized in terms of some fuzzy set-theoretical framework \citep[e.g.,][]{bezdek05}.
\item \textit{One cluster or a few clusters versus a comprehensive clustering?} Thirdly, some clustering methods (e.g., certain methods stemming from computer science) aim at retrieving a single cluster that is in itself optimal in some sense. As a variant, some of these methods subsequently look for the second (third, fourth, etc.) best cluster in itself (possibly after putting aside its previously detected predecessor(s)). A common alternative is to look for a (more or less comprehensive) set of clusters, that is to say, a clustering, which, as a whole, is optimal in one sense or another. An important aspect at this point pertains to the number of clusters in the clustering of interest, which can be pre-specified or chosen during or after the data-analytic process in the narrow sense.
\item \textit{Are all elements to be clustered?} If one is looking for a clustering, which, as a whole, is optimal in some sense, a further relevant question is the following: Should every element of the set of elements that is to be clustered belong to at least one cluster?
\item \textit{Is cluster overlap allowed, and, if yes, is this overlap restricted to nested clusters?} If multiple clusters are looked for, one may wonder whether elements are allowed to belong to multiple clusters, that is, whether (extensional) overlap between clusters is allowed. If the answer to this question is affirmative, a follow-up question reads whether the overlap should (vs. should not) be restricted to nested clusters, meaning that for every pair of overlapping clusters it holds that one of the clusters is a subset of the other. Some authors call the latter type of clusterings hierarchical \citep[though it should be mentioned that both more and less restrictive definitions of a hierarchical clustering have been suggested -- see, e.g.,][]{jardine71, sneath73, vanmechelen97}.
\end{enumerate}

	To typify the organizing principles on the level of the clusters' \textbf{intension}, one may consider both the within- and the between-cluster organization. Regarding the \textit{within-cluster organization}, a first question is what should be the unifying or common ground for elements to belong to the same cluster. Possible answers to this include a common pattern of values for some subset of the variables (e.g., ``$X_3=$ 'present' and $X_5<20$''), small within-cluster dissimilarities, and large similarities of elements to the centroid(s) of the cluster(s) to which they belong. A second question could refer to constraints on the form of within-cluster heterogeneity, such as topological or geometric constraints (e.g., connectedness or convexity), and constraints on the within-cluster (in)dependence structure of variables. Regarding the \textit{between-cluster organization}, a first question may read what should be the discriminating ground for elements to belong to different clusters. Possible answers at this point include large between-cluster dissimilarities and (linear or other) separability. A second question could pertain to constraints on the form of between-cluster differences, such as the requirement that all cluster centroids should lie in a low-dimensional space. \\

\noindent\textit{The criterion or objective function to be optimized} \\

	Any clustering method involves some optimization. The nature of the criterion that is optimized, however, may vary fairly considerably between methods. A critical question at this point is whether or not the optimization concerns an overall objective function. Indeed, a number of clustering procedures proceed in a stepwise fashion, with some specific criterion being optimized in each step, yet without the whole of all steps targeting the optimization of an overall objective function. For many other methods, however, an overall objective function is in place (leaving aside whether its optimization is taken care of by means of a single step or in a stepwise fashion). In the latter case, a follow-up question reads whether the function refers to some kind of goodness of fit (as in, e.g., a likelihood or a least squares fit-related loss function). \\
		
\noindent\textit{The algorithm or computational procedure} \\

	For the actual optimization, clustering methods rely on computing routines. In absence of an overall objective function, these are mere computational procedures. In presence of such a function, they can be considered algorithms. However, in almost all cases, there is no guarantee that the algorithms in question will yield the globally optimal solutions looked for. Major reasons for this are that the optimization problems at hand are typically nonconvex and often (at least partially) discrete in nature, with a ubiquitous presence of local optima \citep[see, e.g.,][]{steinley03}.

\subsubsection{Level 3 - Post-processing of the output:}\label{sec:level3}
A final possible step in the data-analytic process concerns a post-processing of the output of the chosen clustering method. Post-processing often concerns some form of model selection, including the challenging choice of the number of clusters in the final model (insofar as this has not been taken care of during the data analysis in the narrow sense). This could come down to a classical selection among different models (possibly of varying complexities), but also to a selection of one part within one and the same model. As an example of the latter, one may think of the selection of a single partition within a hierarchical clustering, conceived as a collection of nested partitions. Other forms of post-processing imply some kind of simplification of the output of the clustering method. As an example, one may think of applying Bayes' rule to the estimated posterior cluster membership probabilities in mixture models, to convert them into 0/1 membership values. 

\subsection{Data to be used for benchmarking}\label{sec:data}
In the present paper, we focus on dealing with benchmarking questions by means of analyses of data sets. Different types of such data sets can be distinguished:

	A first type is data that are obtained from Monte Carlo simulations. In cases where such simulations have been set up on the basis of a series of a priori, technical or artificial assumptions, we will further call them synthetic simulations (in contrast with the realistic simulations that will be described below).

	A second type is that of empirical data (stemming from one or more substantive domains of interest). Such data may or may not include ``external'' true group information. On the one hand, data that include such information imply an easy possible validation criterion for clustering methods, albeit not necessarily the best or the only legitimate one (see further \ref{sec:evaluationissues}). On the other hand, data that do not include such information are more representative of real clustering problems in which groups are not known in advance and are to be induced from the data. Interestingly, empirical data sets may be conceived as elements sampled from some population of data sets \citep{boulesteix15a}. Challenging questions at this point include how the latter population can be characterized, whether the sample at hand can be considered representative, and which sampling mechanism can be assumed to be in place. Furthermore, when dealing with two or more empirical data sets, one may wonder whether these can be considered exchangeable members of some class of data sets (for which, e.g., the nature of the clusters aimed at would be the same). Obviously, whether or not different data sets are considered exchangeable depends on which distinctive data set characteristics the researcher cares about, and, hence, is ``in the eye of the beholder''.

	Three variants of using empirical data sets in benchmarking studies can further be considered. A first variant is that the data sets are used as given, without any modification. A second variant is where the empirical data sets are modified, for example, by adding outliers or noise variables, or by removing objects or variables. As a third variant, empirical data may be used as input for some kind of realistic simulations (to be distinguished from the synthetic simulations discussed above), which may involve the use of structures, parameter settings or distributions found in analyses of empirical data in the simulations in question. Examples include informing simulations to make the range of the studied factors realistic, and creating simulated data sets by the application of some form of nonparametric bootstrap procedure to a distribution of residuals (i.e., by sampling with replacement from the residuals that result from an analysis of empirical data).

\subsection{Quality criteria for benchmarking}\label{sec:quality}

For the comparative evaluation of methods that is inherently implied by any benchmarking study, numerous quality criteria can be considered. Below, we will outline a few conceptual distinctions between possible criteria, with the primary distinction being between omnibus formal or technical criterion types (which may be used for the evaluation of a broad range of methods) versus criteria that relate to the particular goal of the data analysis under study.

\subsubsection{Omnibus formal / technical criteria: }\label{sec:omnibus}

\noindent\textit{Optimization performance} \\

An obvious first type of criteria relates to the value of the objective or loss function that is optimized in the data analysis. Of course, this presupposes: (a) the presence of an overall optimization criterion, and (b) that all to be compared methods intend to optimize this criterion (without excluding methods that do so in an indirect way only).\\

\noindent\textit{Stability and replicability} \\

A second type of criteria relates to the stability or replicability of the output of the cluster analysis (in terms of its extensional and/or intensional aspects). Stability may be investigated with regard to several choices on the level of the data pre-processing. These may include the entry/removal of outliers, the selection of a subset of the experimental units or variables under study, the type of transformation of variables or data blocks, and the type of conversion of the data structure (e.g., the choice of a dissimilarity measure). Alternatively, stability may also be investigated with regard to choices on the level of the actual data analysis (including algorithm initialization and the values of all kinds of tuning parameters), and the post-processing of the data-analytic output. \\

\noindent\textit{Computational cost} \\

A third type of criteria pertains to the computational cost implied by the data analysis. This type of criteria may be expressed in terms of some measure of a number of operations, computational complexity, or computing time.

\subsubsection{Criteria related to the goal of the data analysis:}\label{sec:criteria}

A second group of criteria refers to the specific primary goal of the data analysis. Subsequently we will consider the case in which this goal concerns the identification of some unknown clustering, and the case of other primary goals. \\

\noindent\textit{Primary goal is identification of clustering} \\

Within this subsection we draw a distinction between three types of criteria: (1) recovery performance, (2) other external validation, and (3) internal validation.
\begin{enumerate}
\item \textit{Recovery performance.} This type of criteria concerns recovery of some aspect of the truth underlying the data. Within a clustering context, relevant aspects may include the true number of clusters, true cluster membership, true cluster characteristics (such as the true values of the cluster centroids or density modes), and the true relevant variables for the clustering. Obviously, this type of criterion requires knowledge about (external) validation information on the truth. 
\item \textit{Other external validation.} Other external validation criteria may, for instance, be relevant if the goal underlying the desire for identifying a clustering is not ``carving nature at its joints'', but is more pragmatic in nature. In the latter case, one may consider external criterion measures that capture this pragmatic goal. As an example, one may think of measures of response to treatment as external validation tools for a patient clustering induced from patient by biomarker data.
\item \textit{Internal validation.} Internal validation criteria may reflect what constitutes a ``good'' or ``desirable'' clustering, both in terms of set-theoretical characteristics of the clusters' extension and in terms of organizing principles on the level of the clusters' intension (see Section \ref{sec:level1}). As examples, one may think of degree of extensional overlap, small within-cluster dissimilarities, and degree of between-cluster separability.
\end{enumerate}

\noindent\textit{Other primary goal} \\

In cases where the induction of a clustering goes with another goal (such as prediction, dimension reduction, or process modeling), with that other goal being the primary concern behind the data analysis, the use of criterion measures that capture this other goal is self-evident. Examples one may think of include measures of predictive quality, and (internal or external) validation measures of assumed underlying processes.

\subsection{Possible answers to benchmarking questions}\label{sec:possible}

Benchmarking questions imply by definition a comparative evaluation of different methods in terms of their performance (without detracting from absolute performance aspects). Answers to such questions on the basis of some benchmarking study may or may not refer to differences between conditions in that study, with conditions referring to subsets of the data types and/or performance criterion variables included in it. Answers that do not refer to such differences may be called ``unconditional''. Examples of unconditional answers include statements about performance rank orders of methods in the study or about an overall winner in the study. Answers that do refer to such differences may be called ``conditional''. Examples of conditional answers include statements such as: ``Method A outperformed method B for such and such data types and/or such and such performance criterion types, whereas for other data and/or criterion types this rank order does not hold or is even reversed''. Importantly, the differences referred to above are limited to within-study differences, and are to be distinguished from differences between results of the study in question and potential results of other studies that involve sets of methods, data and/or criterion types outside the scope of the present study.

 Three variants of unconditional statements (e.g., ``method A outperformed method B'') may further be distinguished. The first (and most demanding) variant reads that the performance rank order of methods A and B holds for each individual combination of a (simulated or empirical) data set and a criterion variable in the study. The second and the third variant presuppose simulated benchmark data that have been generated according to some design in which a number of data characteristics have been systematically varied. The second variant then reads that the rank order holds for all combinations of data types (in the simulation design) and criterion variables after averaging across all data sets within each data type (i.e., after averaging across all (simulated) replications within each cell of the design). Finally, the third variant reads that the rank order holds on average across all data sets and all criterion variables under study. That being said, one may wonder how at this point a proper averaging should be conducted. For an averaging across different data types, this may require a suitable theoretical framework for the population of data types of interest. With regard to different criterion measures, averaging looks even more challenging given obvious commensurability problems related to different criterion measures. Apart from that, unfortunately, ``convenience averaging'', just ignoring the raised issues, is ubiquitous.
 
For their part, conditional answers imply that concepts such as global optimality are to be replaced by more shaded concepts such as Pareto optimality (with a method being Pareto optimal if no other method exists that outperforms the first for some combination of a data type and a criterion variable without leading to a worse performance for some other combination of a data type and a criterion variable). Furthermore, if for a given criterion measure the nature of the optimal method differs across data types, one ends up in a situation that, formally speaking, is very similar to typical situations in clinical trial research in which there is no therapeutic alternative that outperforms all other alternatives for all (types of) patients, and which call for a precision medicine approach. Likewise, within a benchmarking context, one may wish to look for so-called optimal data-analytic regimes, that is to say, comprehensive decision rules that specify, given some criterion measure, for which data types which methods are to be preferred  \citep{boulesteix17a, doove17}. 

\section{How to address benchmarking questions in clustering: Practicalities and foundational guidelines}\label{sec:practicalities}
\subsection{Choice of methods to be compared}\label{sec:choice}
\subsubsection{Issues:}\label{sec:choiceissues}

By definition, benchmarking implies a comparative evaluation of different methods. Hence, for any benchmarking study in clustering, two or more methods have to be chosen, with the methods in question differing from one another with regard to at least one of the three steps of the data-analytic process: (1) the input preprocessing, (2) the data analysis in the narrow sense, and (3) the output post-processing. Moreover, apart from choices on general method aspects, quite a few specific aspects have to be decided on as well, including the choice of an implementation, an initialization, tuning parameters, convergence criteria, and time limits.

\subsubsection{Recommendations:}\label{sec:choicerecommendations}
	We propose three sets of recommendations:
\begin{itemize}
\item	Make a suitable choice of competing methods. At this point ``suitable'' means that:
\begin{itemize}
\item	The intended scope of the methods should be well-defined, well-justified from the point of view of the aims of the clustering, and well-reported. Moreover, the intended scope should be covered properly. For example, within the intended scope, the choice should be sufficiently broad and should not be limited to close neighbors or variants of the same method (unless the intended scope would be limited to such variants indeed).
\item	Do include known strong competitors, that is to say, methods that have been shown in previous studies to outperform others for the type(s) of problems and criteria under study \citep{boulesteix15}. Otherwise, in the search for strong competitors, one should take into account that clustering methods are being developed and studied in a broad range of research domains (including discrete mathematics, statistics, data analysis, computer science, and bioinformatics), and therefore not to limit this search to a single research domain only.
\item If methods are to be compared that differ from one another in terms of different aspects (e.g., initialization method and type of iteration scheme, or type of variable selection and type of algorithm), one may consider (if meaningful and possible) to orthogonally manipulate these different aspects according to a full factorial design (to enhance inferential capacity in terms of disentangling the effects of each of these aspects and their interactions).
\end{itemize}
\item	Make full information on the competing methods available, in view of reproducibility \citep{donoho10, hofner16, peng11} and of enabling follow-up research. This means:
\begin{itemize}
\item	Make the code of the methods under study fully available.
\item	Disclose full information on aspects such as initialization, values of tuning parameters, convergence criteria, time limits, random number generators and random number seeds.
\end{itemize}
\item	Organize a fair comparison in terms of the choices of the specific aspects (with fairness meaning that certain methods are neither wrongfully favored nor put at a disadvantage):
\begin{itemize}
\item	This implies: (1) that the same amount of a priori information is made available to the different methods; (2) fair choices with regard to the specification of tuning parameters (such as the number of starts); (3) fairness in terms of the number of operations/amount of computing time apportioned to the different methods under study (with the obvious problem that the latter may depend fairly strongly on the particular implementation that has been chosen).
\item	This presupposes that the same amount of time is spent to implement the different methods, to handle bugs, to choose tuning parameters, and so on.
\end{itemize}
\end{itemize}

\subsection{Data sets used for benchmarking}\label{sec:data}
\subsubsection{Issues:}\label{sec:dataissues}
In case of data that are obtained from synthetic simulations, choices have to be made regarding different aspects. A first of these pertains to the factors that are to be manipulated in the simulations (along with the levels of each chosen factor). Examples include the number of objects, the number of variables, the proportion of variables that are relevant for the clustering, the number of clusters, (in)equality of cluster sizes, the cluster generative distribution, the amount of distributional overlap, etc. A second aspect pertains to the experimental design of the simulation setup. The most simple option at this point is a full factorial design. A third aspect pertains to the number of replications in each cell of the design. For the actual simulations, one may either rely on own code or on existing data generators. Regarding the latter, over the past decades quite a few generators have been proposed, including the \citet{milligan85}  algorithm for generating artificial test clusters, OCLUS \citep{steinley05}, the \citet{qiu06} random cluster generation algorithm, and MixSim \citep{melnykov12}. In the justification of these generators quite some emphasis has been put on the aspects of separability and overlap; in all cases, however, overlap refers to \textit{intensional} rather than to extensional overlap, that is to say, overlap in terms of variables or component distributions, with all generated clusterings being partitions. Advantages of existing data generators include their ease of use, and comparability with results of other studies that have used the same generators; disadvantages include that they produce a somewhat limited scope of data sets. A blueprint of a novel device for simulating data for benchmarking in unsupervised learning has been designed by \citet{dangl}. This blueprint comprises the plan of a web repository and an accompanying \texttt{R} \citep{r17} package. The web repository will host a collection of benchmarking setup files. These will be script files that may generate metadata objects, which contain cluster-specific parameter information, information on the data-generating function, random seeds, the type and version of the random number generator, etc. One will be able to use the accompanying \texttt{R} package for the actual production of metadata objects and for the subsequent generation of data sets on a local computer. The whole framework will allow, on the one hand, users to upload script files that include metadata information on previously used benchmarking data sets, and, on the other hand, other users to download the script files to exactly reproduce the previously used data sets or to generate new data sets by modifying the corresponding metadata information. Key advantages of the framework will include that it will make the generation of simulated benchmarking data sets fully transparent and reproducible. 

	Alternatively, one may wish to use a collection of empirical data sets in benchmarking studies (which may be considered a sample from the population of all empirical data sets). One should take into account that, within such a sample, typically empirical correlations between data characteristics show up (unlike in simulated setups in which data characteristics can usually be orthogonally manipulated). Such correlations can possibly hamper the inferences one may draw from benchmarking studies using the data in question. Two remarks on empirical data can further be made. First, to retrieve empirical data sets one could rely on various repositories. A candidate in this regard is the Cluster Benchmark Data Repository that has been recently launched by the authors of the present paper within the Cluster Benchmarking Task Force of the International Federation of Classification Societies (IFCS) at \href{http://ifcs.boku.ac.at/repository/}{http://ifcs.boku.ac.at/repository/}. This repository intends to provide a wide variety of well-documented, high quality data sets (with and without given ``true'' clusterings) for use in benchmarking in cluster analysis. A unique feature of the repository is that each data set is supplied with comprehensive meta-data, including documentation on the specific nature of the clustering problem and on characteristics that useful clusters should fulfill (with scientific justification). Secondly, when drawing a sample from the population of all empirical data sets, one could follow either a top-down or a bottom-up approach. In a top-down approach, one may start from a profile of data characteristics (along with common criteria of what would constitute a ``good'' clustering), to subsequently look for a collection of data sets that sufficiently covers this profile (with the profile in question being formally analogous to the inclusion criteria that patients have to fulfill to be recruited in a clinical trial -- see also \cite{boulesteix17b}). Alternatively, one could take a bottom-up approach, which comes down to starting from a single data set (and an in-depth analysis of what would constitute a ``good'' clustering for it), to subsequently look (e.g., in a repository under study) for a collection of data sets with similar profiles.

\subsubsection{Recommendations:}\label{sec:datarecommendations}
	We propose three groups of recommendations:
\begin{itemize}
\item	Make a suitable choice of data sets and give an explicit justification of the choices made. This means that:
\begin{itemize}
\item	the intended scope of generalization for the data sets should be well-defined, well-justified, and well-reported;
\item	the field/population of data sets of interest should be well covered \citep{boulesteix13};
\item	the data sets should be sufficient in number to allow for reliable inferences (also taking into account power issues: see further Section \ref{sec:analysis}) \citep{boulesteix15};
\item	whenever feasible, one should consider to carefully combine synthetic simulations and empirical data as these may yield complementary information;
\item	if simulated data sets are to be analyzed that differ from one another in terms of different aspects (e.g., sample size, degree of cluster separability), one should consider (if meaningful and possible) to orthogonally manipulate these different aspects according to a full factorial design (to enhance inferential capacity);
\item	whereas, when proposing a new method, one may typically wish to show that this method does something useful on some data sets that is not covered by already existing methods, one should explicitly and honestly report about how the data sets in question were selected (e.g., because they yielded favorable results for the newly proposed method) \citep{boulesteix17b}; moreover, when proposing a new method, the inclusion of cases in which this method does not work (i.e., foils) may be particularly informative to clarify the method's actual scope.
\end{itemize}
\item	In case of simulated data, organize a fair comparison in terms of the relation between the methods under study and the data-generating mechanisms of the simulation study, with fair meaning that these mechanisms should not unilaterally favor some of the methods (because the methods explicitly or implicitly assume these mechanisms to be in place).
\item	Disclose full information on the data sets that are used in view of reproducibility \citep{dangl, donoho10, hofner16, peng11} and of enabling follow-up research. This means that:
\begin{itemize}
\item	for simulated data sets implementable data-generating code with full information on cluster-specific parameters, the data-generating function, random seeds, the type and version of the random number generator, and so on should be provided;
\item	for empirical data sets the full data sets should be provided, with sufficient detail on format, codes used to denote missing values, preprocessing, and so on. 
\end{itemize}
\end{itemize}

\subsection{Evaluation measures}\label{sec:evaluation}
\subsubsection{Issues: }\label{sec:evaluationissues}
	On the level of \textit{formal/technical criterion measures}, indices of optimization performance are immediately implied by the objective function that is optimized in the data analysis. Examples include the trace of the pooled within-cluster variance-covariance matrix W that is minimized in $k$-means \citep{macqueen67}, as well as least squares and likelihood objective functions. Specific challenges that go with these measures include that their values are often not comparable across different benchmarking setups or data sets, and that one usually does not know the global optimum for them for a data set under study. \citep[In case of comparability problems, one may consider trying to solve them by means of some kind of normalization procedure: see, e.g., ][]{brusco07}. For their part, regarding formal criterion measures referring to stability and replicability, a broad range of (dis)similarity measures could be considered \citep[see, e.g., ][]{breckenridge89}.

	On the level of \textit{criterion measures related to the goal of the data analysis}, we first consider the case in which the primary goal is the identification of a clustering:
\begin{itemize}
\item	Various indices of recovery performance may be considered. Examples include the Adjusted Rand Index \citep{hubert85, steinley16} to measure cluster membership recovery in a partitioning context, the mean squared difference between an overlapping cluster membership matrix or some parameter vector/matrix and its true counterpart (minimized across all possible permutations of the clusters), and measures of recall and precision to measure recovery of the subset of variables that is truly relevant for the clustering of interest \citep{steinley08}. Importantly, even if data go with true grouping information, this is not necessarily the best or only legitimate validation criterion, as in cluster analysis the ``truth'' may depend on the context and the goals of the analysis \citep{hennig15}, and therefore reasonable clusterings may or may not correspond with the given grouping information.
\item	Criterion measures that involve external validation information other than information on the truth may include effect sizes (such as $R^2$- and $\eta^2$-type measures) associated with regression analyses, analyses of variance, or discriminant analyses that relate cluster membership to one or more external validation variables.
\item	Alternatively, one may wish to consider criterion measures of different aspects of internal validation such as within-cluster homogeneity, between-cluster separation, and similarity of cluster members to their cluster centroid \citep{hennig15, hennig15a, milligan81}. A broad range of measures may be considered for this purpose. Examples include the Average Silhouette Width \citep{rousseeuw87} or a fuzzy extension of it \citep{campello06}, which in a partitioning context measures how similar objects are to their own cluster compared to other clusters, and the cophenetic correlation coefficient \citep{sokal62}, which in a hierarchical clustering context measures the relation between the ultrametric implied the hierarchical clustering and the proximity data from which this clustering was derived.
\end{itemize}
	Secondly, we may consider the case with a primary goal other than the identification of a clustering. As examples of measures one may think of the proportion of explained criterion variance associated with prediction models and of goodness-of-fit measures of process models that involve some kind of latent heterogeneity that is captured by a clustering.

\subsubsection{Recommendations: }\label{sec:evaluationrecommendations}
\begin{itemize}	
\item Think very carefully about the nature of the quality criterion/criteria that you may wish to use, not in the least because different quality criteria may imply different optimal clustering solutions \citep{hennig15, hennig15a}. Make your choice explicit along with a justification of it. Consider multiple criteria if appropriate, optionally with an index of their relative importance (similar to the distinction between primary and secondary endpoints in clinical trial research).
\item	Make a suitable choice of criterion measures in an a priori way (i.e., before any comparative data analysis). At this point suitable means that: 
\begin{itemize}
\item	the measures should capture the chosen target criterion/criteria;
\item	the application of the measures in question in the context under study should be technically correct (e.g., the use of the Adjusted Rand Index as a measure of cluster recovery only makes sense within the context of partitions);
\item	the values of the chosen measures should be (made) comparable across all benchmarking data sets.
\end{itemize}
\item	Examine the interrelations between the performance regarding different criterion measures whenever appropriate. For example, it may be desirable to examine the relation between recovery of the true number of clusters and recovery of the true cluster memberships, with recovery of the latter possibly being better in applications with a wrong number of clusters.
\end{itemize}

\subsection{ Analysis, results, and discussion}\label{sec:analysis}
\subsubsection{Issues: }\label{sec:analysisissues}
	One may wish to look for \textit{unconditional conclusions} on the performance rank order of the methods included in a benchmarking study, in terms of: (1) a consistent rank order across all combinations of individual data sets and criterion variables included in that study, or (2) a consistent rank order across all data types and performance criterion variables, after averaging across all data sets (or simulated replications) within each data type, or (3) a rank order on an aggregate level. Regarding the third option, insofar as criterion values are comparable across data sets and insofar as only a single criterion is focused on (or multiple commensurable criteria), one may compare the methods under study in terms of their average performance (averaged across data sets and possibly different criteria). Optionally, one may further wish to test the null hypothesis of no differences, which may be taken care of by means of testing the main effect of the (within-subject) method factor in a (possibly factorial) analysis of variance (in which data sets act as experimental units). Alternatively, in case of comparability or commensurability obstacles, one may wish to make up some type of consensus ranking of the methods (with consensus referring to an aggregation across data types and/or criterion variables), for example, based on Kemeny's axiomatic framework \citep[see also][]{eugster11}. Notice that in principle all of the above could also be applied in case of empirical benchmarking data sets \citep{boulesteix15a}. Obviously, in the latter case critical ANOVA assumptions such as random sampling from a population of real data sets of interest and independent and identically distributed data can be expected to be violated (with, e.g., the sampling process most probably being subjected to several kinds of selection bias); nevertheless, without detracting from the desirability to acknowledge these likely violations and from the need for caution they imply, classical power calculations may be meaningful in benchmarking studies based on real data sets, as they may provide a useful order of magnitude for the required sample sizes and inferential error rates \citep{boulesteix15a}.

	In the absence of comparability and commensurability problems, \textit{conditional conclusions} may be drawn from a benchmarking study by investigating method by data generation and/or criterion factor interactions in (repeated measures) analyses of variance. In the presence of such problems, one may look at the interactions in question in a more informal, descriptive way; alternatively, one may then also wish to identify the set of Pareto-optimal methods. Importantly, regarding the interactions in question, one should take into account that the testing of them in a reliable way typically requires larger sample sizes than the testing of main effects. This may be even more notable in the case of benchmarking with empirical data sets because of to be expected nonnegligible correlations between data characteristics.
	
	Finally, all types of conclusions one may draw from a benchmarking study always only hold conditional on the scope of that study in terms of the set of methods as well as the set of data types and performance criterion variables included in it. Moreover, whereas within a study with a limited number of data sets, data types and criterion variables a consistent rank order of performance may show up, universal winning methods across all possible data sets, data types and criterion variables can be shown not to exist, as for a given data set different quality criteria may imply different optimal clustering solutions \citep{hennig15, hennig15a}.

\subsubsection{Recommendations: }\label{sec:analysisrecommendations}
\begin{itemize}
\item	Whereas unconditional statements about performance rank orders of methods may be useful summaries:
\begin{itemize}
\item	We recommend to investigate always explicitly whether the results of one's study are subject to sizeable within-study differences between conditions.
\item	Whenever unconditional statements are reported, it should be made very clear whether or not they are based on some kind of averaging, and, if yes, on which one.
\item	As means can be dominated by a few situations in which methods do very badly, one may wish to consider (next to averages) more comprehensive representations of the distribution of results (e.g., by means of boxplots).
\end{itemize}
\item	If possible, examine main and interaction effects by means of a (repeated measures) factorial analysis of variance, given its obvious inferential advantages. When doing so, however, take care of an adequate reporting of effects:
\begin{itemize}
\item	Go beyond significance levels (which are often meaningless because of large sample sizes in many simulation studies), report effect sizes (of, e.g., $\eta^2$-type), and consider to use a threshold on them to discuss only substantial effects.
\item	Go beyond main effects and also look at interactions.
\item	Inspect the contents of the found effects, test contrasts whenever needed to clarify the patterns involved in them, and provide an insightful reporting of these patterns making use of custom-made graphical displays.
\end{itemize}
\item	Properly address developers of methods:
\begin{itemize}
\item	Provide insight into the why of the found effects (in terms of underlying mechanisms, relations between characteristics of methods and data, etc.).
\end{itemize}
\item	Properly address practitioners:
\begin{itemize}
\item	Explain which method gives the best results for which cluster concepts, for which criteria, and for which types of data. At this point, qualitative or disordinal method by data characteristic interactions should be highlighted insofar they imply that the best method differs across criteria and/or data types. Otherwise, if the latter type of interactions would show up, one may further consider to look for an optimal data-analytic regime, that is to say, a decision rule (e.g., represented by a decision tree) that specifies, given some criterion measure, for which data types which methods are to be preferred \citep{doove17}.
\item	From a pragmatic viewpoint, one should focus on recommendations in terms of data characteristics that are known to the researcher outside the context of Monte Carlo simulations (such as, e.g., data size); in case other characteristics (such as, e.g., the true number of clusters) appear to play a critical role in disordinal method by data characteristic interactions, one may wish to try to replace them by proxies that are known in data-analytic practice \citep{doove17}.
\end{itemize}
\item	Do not overgeneralize the obtained results. Never ever forget to mention very explicitly in the final conclusion the restricted scope/area of the problem that has been looked at in the study (in terms of selected methods, (types of) data sets, and the specific performance criteria being used), and always explicitly discuss the limitations of this scope.
\end{itemize}

\subsection{Authorship}\label{sec:author}
\subsubsection{Issues: }\label{sec:authorissues}
	The relationship between the author(s) of the benchmarking study and the authorship of the methods investigated in this study deserves special attention. Obviously, authors who propose new methods are in practice typically obliged to show that these methods outperform competitors in at least some situations, as a kind of existential justification of these methods. This, however, may also be looked at as an instance of publication bias, in that it is difficult to publish a paper on a new method that does not outperform competitors \citep{boulesteix15b}. Moreover, in practice, the need of evidence to justify the existence of a newly proposed method may degenerate into some kind of less warranted over-optimism (e.g., in terms of reports of benchmarking evidence that that method uniformly outperforms a number of competitors for all criteria and all data types under study). This is further evidenced by a survey conducted by \citet{boulesteix13a}  of articles published in seven high-ranked computational science journals; from this survey it appeared that benchmarking studies that were part of a paper in which new methods were proposed very often identify the new methods as winner. Moreover, not surprisingly, papers suggesting new methods appeared to never be negative with respect to the new methods.

\subsubsection{Recommendations: }\label{sec:authorrecommendations}
\begin{itemize}
\item	It is important to disclose in benchmarking studies any possible conflicts of interest (e.g., vested interest in some of the methods under study).
\item	``Neutral'' comparison studies by authors without vested interests and, for the authorship as a collective, ideally with approximately the same level of familiarity with all methods under study \citep{boulesteix17b} should be especially encouraged. This further implies that journals should welcome reports of them and should guard against any possible publication biases against neutral comparison studies of existing methods \citep{boulesteix13a}. In cases where foils were missing in earlier work, it would be helpful if neutral comparison studies could contribute some of these.
\end{itemize}

\section{A few illustrative examples}\label{sec:examples}
	In this section we will briefly discuss a few illustrative examples of benchmarking studies in the clustering area. Rather than providing self-contained summaries of the studies in question, we will highlight a few distinctive features of each of them.

\subsection{Steinley (2003)}\label{sec:steinley}
	\citet{steinley03} presented an impactful comparison of three widespread commercial software packages and one multistart MATLAB routine for $k$-means analysis, one of the most basic and prototypical methods for cluster analysis. In this study, he addressed a classical problem in the optimization of overall objective or loss functions in many methods of cluster analysis, that is to say, the ubiquitous occurrence of locally optimal solutions. For this purpose he made use of two empirical data sets and simulated data (on the basis of a full factorial design). Criterion variables included the value of the loss function (for the empirical data sets) and recovery of the true clustering (for the simulated data sets). The comparison suggested that the commercial packages most often return locally optimal solutions or solutions that imply a worse recovery of the underlying true clustering than those produced by a suitable multistart procedure.

\subsection{Steinley \& Brusco (2008)}\label{sec:brusco}
	\citet{steinley08} set up an extensive comparison of eight different variable selection procedures for model-based and non-model-based clustering. For this purpose they made use of a full factorial design with 6,804 conditions and three replications per cell, which resulted in 20,412 simulated data sets. Subsequently, they applied all eight methods under study to each data set, which resulted in 163,296 clustering runs.

\subsection{Anderlucci \& Hennig (2014)}\label{sec:hennig}
	\citet{anderlucci14} compared two approaches for clustering categorical data that are based on totally different conceptual frameworks: partitioning around medoids \citep{kaufman90}, which attempts to produce homogeneous clusters with low within-cluster dissimilarities, and latent class modeling \citep{vermunt02}, which tries to estimate an underlying mixture model. In the study (which also used a full factorial design), two different evaluation criteria were used, one corresponding to each approach: the average silhouette width for finding clusters that are homogeneous and well-separated from each other in terms of dissimilarities, and the adjusted Rand index to measure the recovery of the clusters that were true according to the underlying mixture model. One result was that the latent class approach was in most situations surprisingly competitive regarding the average silhouette width.

\subsection{Schepers \& Van Mechelen (2006)}\label{sec:schepers}
	\citet{schepers06} set up a four-part benchmarking study to investigate the sensitivity of different partitioning algorithms to local optima. Interestingly, the rank order of the algorithms found in a study with synthetically simulated data sets (Part 1) did not correspond to the rank order of the same algorithms when applied to empirical data (Part 2). Parametric bootstrap tests (Part 3) revealed that assumptions of the stochastic model underlying the data-generating mechanism used in Part 1 were violated in the empirical data of Part 2. In a new study with simulated data using a modified data-generating mechanism (Part 4) this hypothesis was confirmed.

\section{Concluding remarks}\label{sec:conclusion}
	The obvious primary take home recommendation to investigators of clustering methods is to invest in benchmarking. This should not only be an absolute requisite when proposing new methods, but should likewise be taken care of in follow up research, not in the least also by authors who were not involved in the development of the methods under study.

	If we further analyze the specific recommendations listed in the present paper, we may conclude that many of them are relevant for benchmarking in statistics and data analysis in general. Examples include:
\begin{itemize}
\item	Do include in any benchmark attempt known strong competitors for the target methods under study.
\item	Orthogonally manipulate aspects of methods and characteristics of simulated benchmark data insofar possible.
\item	Make full information available on methods (code, values of tuning parameters etc.) and benchmark data in view of full reproducibility.
\item	Make use of a suitable combination of simulations and empirical data when creating a collection of benchmark data sets.
\item	Take care of fairness when setting up benchmark comparisons.
\item	Always examine whether results of a benchmarking  study are subject to sizeable within-study differences between data types and/or criterion variables.
\item	Disclose any possible conflicts of interest.
\item	Journals should especially welcome ``neutral'' comparisons by authors without vested interests in any of the methods under study.
\end{itemize}
	Apart from this, we also listed several specific recommendations that are more typical for the clustering domain. Perhaps the most fundamental of these is to go for a deep reflection of what constitutes, within a particular context and given particular research questions and aims, a good or desirable clustering. The result of this may have far-reaching consequences for the choice of relevant methods, benchmark data sets, and evaluation criteria. Much of this fundamental recommendation may be linked to the lack of an obvious metric for a performance evaluation of methods for unsupervised learning in general and for clustering in particular, which necessitates the deep reflection referred to above. Such a lack could be perceived as a bottleneck. Yet, it could also be perceived as an opportunity (and even as a blessing) that could finally lead to insights that may be inspiring for the clustering domain as well as for other subdomains of statistics and data analysis.

\section{Glossary}\label{sec:glossary}
\begin{itemize}
\item	\textit{benchmarking}: comparative evaluation of two or more data-analytic methods
\item	\textit{cluster analysis}: type of unsupervised statistical learning that results in a grouping of elements in one or more sets that are not known prior to the analysis
\item	\textit{cluster extension}: cluster membership
\item	\textit{cluster intension}: features, attributes, organizing principles, or definition associated with cluster
\item	\textit{conditional answer to benchmarking questions}: answer that explicitly acknowledges that within a particular benchmarking study the (total or partial) performance rank order of methods differs across different subsets of the data types and/or performance criterion variables included in that study
\item	\textit{data-analytic process}: three-step process that comprises: (1) a pre-processing of the data, (2) the data analysis in a narrow sense (i.e., leaving aside the other two steps), and (3) a post-processing of the output of the previous step
\item	\textit{external part of the data}: part of the data that is not to be used in the actual cluster analysis but that can be used for a subsequent validation of the results of the cluster analysis
\item	\textit{internal part of the data}: part of the data that is to be used in the actual cluster analysis
\item	\textit{Pareto optimality}: in a multicriteria choice context, a choice alternative is Pareto optimal if no other alternative exists that outperforms the first on a criterion variable without performing worse on another criterion variable
\item \textit{realistic simulations}: simulations that involve information (structures, parameter settings, \ldots) derived from analyses of empirical data
\item	\textit{synthetic simulations}: simulations set up on the basis of a series of a priori, technical or artificial assumptions
\end{itemize}

\end{document}